\documentclass[aps,pra,twocolumn,floatfix,longbibliography]{revtex4-2}
\usepackage[utf8]{inputenc} 
\usepackage{graphicx}
\usepackage{amsmath}
\usepackage{amsfonts}
\usepackage{mathtools}
\usepackage{float}
\usepackage{tabularx, booktabs}
\usepackage{color}
\usepackage{braket}
\usepackage{bm}
\usepackage{comment}

\usepackage{hyperref}

\hypersetup{
	colorlinks=true,
	linkcolor=blue,
	citecolor=blue,
	filecolor=magenta,
	urlcolor=cyan,
	breaklinks=true
}
\newcommand{\papertitle}{Bose-Einstein condensation on axially-symmetric surfaces}

\DeclareSymbolFont{sfletters}{OML}{cmbrm}{m}{it}

\begin{document}
	
\title{\papertitle}
\author{Andrea Tononi}
\affiliation{ICFO-Institut de Ciencies Fotoniques, The Barcelona Institute of Science and Technology, Castelldefels (Barcelona) 08860, Spain.}
\affiliation{Department de F\'isica, Universitat Polit\`ecnica de Catalunya, Campus Nord B4-B5, E-08034, Barcelona, Spain}

\author{Gregory E. Astrakharchik}
\affiliation{Department de F\'isica, Universitat Polit\`ecnica de Catalunya, Campus Nord B4-B5, E-08034, Barcelona, Spain}

\date{\today}

\begin{abstract}
We investigate the phenomenon of Bose-Einstein condensation in ideal bosonic gases confined to axially-symmetric surfaces of revolution. The single-particle Schrödinger equation is formulated on a general surface and then explicitly solved in the ellipsoidal and toroidal geometries to determine the one-body energy spectrum. We discuss how the curved geometry impacts the quantum statistical properties of ideal Bose gases confined on these surfaces. Specifically, we observe that Bose-Einstein condensation is suppressed when the surface aspect ratio is increased and, correspondingly, it becomes highly elongated and acquires a one-dimensional character. We also evaluate the Bogoliubov excitation spectrum, providing insights into the collective excitations of the condensate. Our results establish the conditions to achieve quantum degeneracy in curved manifolds, thus guiding forthcoming experiments with thin shells, and set the basis for solving the few-to-many body problem in general surfaces of revolution.
\end{abstract}

\maketitle 

\section{Introduction}
Bose-Einstein condensation (BEC) occurs when a macroscopic fraction of the particles of a system occupy the same single-particle state \cite{pitaevskii2016bose}.
While BEC was first implicitly observed in superfluid helium, in the context of ultradilute gases it was experimentally realized in harmonically trapped atoms \cite{anderson1995observation,davis1995bose}. 
Unlike in helium, ultracold gases offer spectacular control over both interactions and confinement geometries, enabling studies in various trapping configurations, including lattices and boxes \cite{PhysRevLett.86.4447, meyrath2005bose}. 
Notably, various experiments have realized the confinement of atomic gases in two-dimensional configurations \cite{hadzibabic2006berezinskii, kruger2007critical}, whose dynamics is restricted to zero-point motion along the strong confinement direction and free otherwise.
Recently, the experimental study of ultracold atoms in two-dimensional curved geometries has become an emerging {research} trend \cite{carollo2022observation,jia2022expansion, guo2022expansion, dubessy2025quantum}.
Analyzing the case of an ideal Bose gas confined to curved surfaces seems particularly intriguing. 
In contrast to three dimensions, where a finite critical temperature exists for Bose-Einstein condensation of an ideal gas, the Mermin-Wagner theorem \cite{PhysRevLett.17.1133} forbids condensation in an infinite two-dimensional plane.
However, Bose-Einstein condensation can still occur in finite-size systems \cite{PhysRevLett.89.280402}, raising an interesting question about how curved geometries influence this phenomenon \cite{tononi2023low}.

On the theoretical side, various studies have focused on the quantum statistics of ultracold atoms confined on spherical and ellipsoidal shells \cite{tononi2024shell, moller2020bose, he2022bcs, he2024two, biral2024bose}.
These investigations pointed out that the curved confinement changes the energy spectrum of the system with respect to the analogous flat counterparts, producing quantitative geometric-dependent corrections to the system thermodynamics \cite{tononi2022scattering, he2022bcs}. 
Other studies have shown, for instance, how the variation of geometric parameters affects the critical Bose-Einstein condensation temperature \cite{tononi2019bose, bereta2019bose, kurt2024shape}.
However, we note that so far no analyses of the Bose-Einstein condensation transition have been conducted for gases confined in some of the simplest purely-two-dimensional geometries, such as tori and ellipsoidal surfaces.
Analyzing this phenomenon would not only guide their experimental realizations \cite{PhysRevA.75.063406, carollo2022observation, guo2022expansion, gentile2025} but also set the basis for the development of few-body physics in new curved geometries.

In this paper we discuss the phenomenon of Bose-Einstein condensation in axially-symmetric surfaces of revolution, elucidating how the curved geometry affects the quantum statistical properties in ellipsoids and tori.
In particular, we first formalize the single-particle Schrödinger equation for generic surfaces of revolution.
Then we focus on the specific cases of ellipsoid and torus, and numerically determine the one-body energy spectrum {and eigenfunctions}.
This result allows us to analyze the Bose-Einstein condensation phenomenon and to determine the Bogoliubov energy spectrum for a gas confined in these manifolds.

Our main result, namely the suppression of Bose-Einstein condensation in elongated geometries, can be observed in experiments with quasi-two-dimensional Bose gases trapped near curved manifolds.
Conveniently, our method can be easily extended to include the eventual trap inhomogeneities of experiments with thin ellipsoidal shells \cite{carollo2022observation}, and can therefore support them towards the goal of reaching the condensate regime.
Our results can also guide the forthcoming realization of Bose-Einstein condensates confined near a toroidal surface \cite{PhysRevA.75.063406, gentile2025}.

\section{Schrödinger equation on axially symmetric surfaces}
\begin{figure}[hbtp]
\centering
\includegraphics[width=0.99\columnwidth]{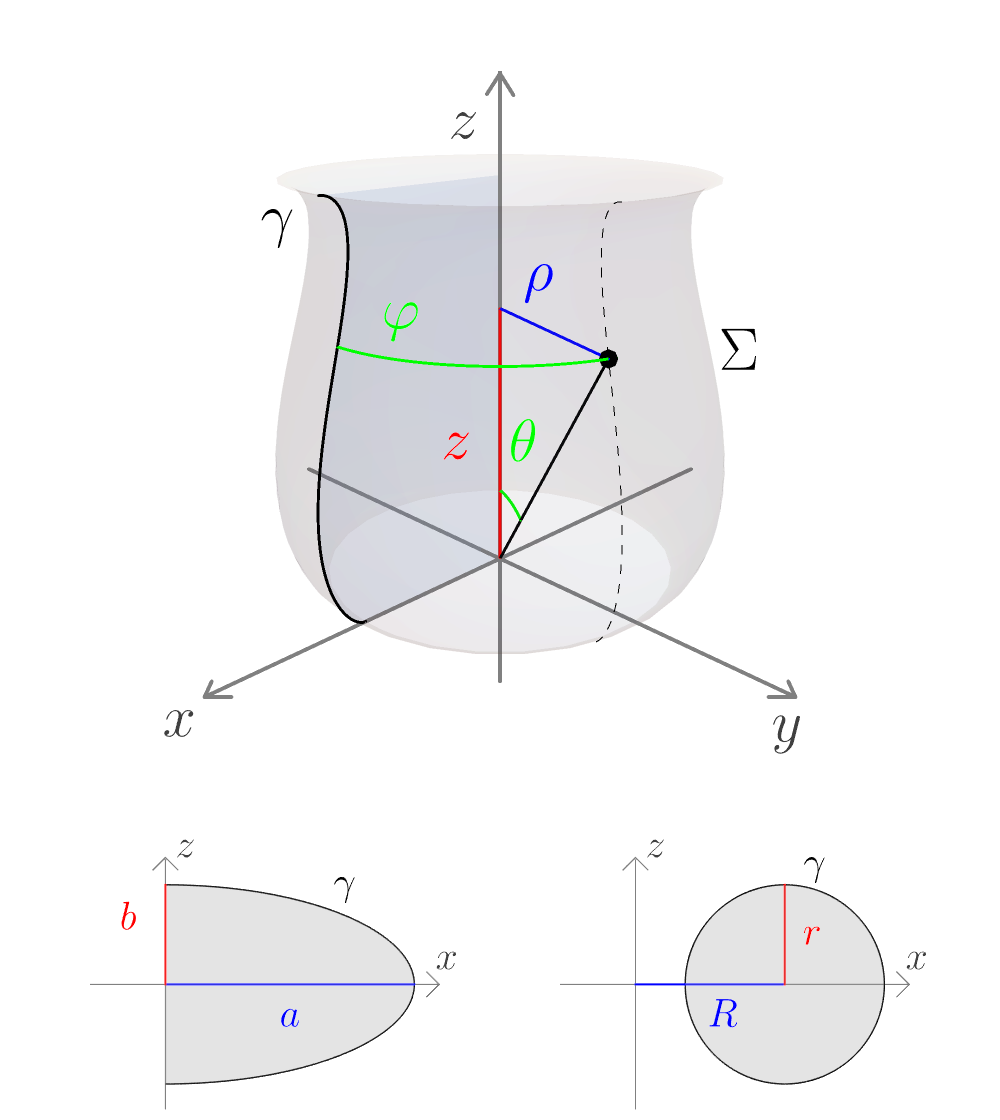}
\caption{ Shown on top is an illustration of the system geometry and the coordinate system parametrizing the axially-symmetric surface $\Sigma$. The surface is obtained from the revolution of the curve $\gamma$ around the $z$ axis.
The bottom shows an illustration of the curves $\gamma$ generating the ellipsoidal surface (left) and the toroidal surface (right) and the respective geometric parameters.
}
\label{fig1}
\end{figure}

A single quantum particle moving on the surface $\Sigma$ satisfies the Schrödinger equation 
\begin{equation}
(\hat{T} - \epsilon)\Psi = 0,
\label{Schreq}
\end{equation}
where $\hat{T}$ denotes the kinetic energy operator restricted to the surface, $\epsilon$ is the energy eigenvalue, and $\Psi$ is the unit-normalized wave function.
We assume that the surface $\Sigma$ is a non-intersecting axially-symmetric manifold, parametrized by 
\begin{equation}
\Sigma = (\rho(\theta) \cos\varphi,\rho(\theta) \sin\varphi,z(\theta)), 
\end{equation}
where $\theta \in I$ parametrizes both the distance $\rho(\theta)$ from the $z$ axis and the $z$ coordinate $z(\theta)$, while $\varphi \in [0,2\pi]$ is the azimuthal angle (see Fig.~\ref{fig1}). 
Note that the surface $\Sigma$ is generated by the revolution of the differentiable curve $\gamma = (\rho(\theta),0,z(\theta))$ along the $z$ axis and its area can be evaluated through the Guldinus theorem as
\begin{equation}
S = \int_{0}^{2\pi} d \varphi \int_I  d\theta \, ||\partial_{\theta}\Sigma \times \partial_{\varphi}\Sigma || = \int_{0}^{2\pi} d \varphi \int_I  d\theta \, \rho(\theta) t(\theta),
\label{surfacearea}
\end{equation}
where $t(\theta) = [\rho'^2(\theta) + z'^2(\theta)]^{1/2}$ is the modulus of the tangent vector to $\gamma$, with the prime symbol denoting the first derivative.

The Schrödinger equation~\eqref{Schreq} in these coordinates, for a particle of mass $M=1$ and setting $\hbar=1$, reads
\begin{align}
\begin{split}
\bigg[\hat{T}_{\theta} + \frac{\hat{L}_z^2}{ 2 \rho^2(\theta)} 
- \epsilon  \bigg] \Psi(\theta,\varphi) = 0,
\label{schreq}
\end{split}
\end{align}
where 
\begin{equation}
\hat{T}_{\theta} = - \frac{1}{2 t^2(\theta)} \left\{ \partial_{\theta}^2 + \left[\frac{\rho'(\theta)}{\rho(\theta)} - \frac{t'(\theta)}{t(\theta)} \right] \partial_{\theta}  \right\}, \ \hat{L}_z^2 = -\partial_{\varphi}^2,
\label{Tgeneric}
\end{equation}
results from directly evaluating $\hat{T}$ in terms of the Laplace-Beltrami operator (see Appendix \ref{appA}).
Due to the rotational symmetry around {the} $z$ axis, the angular momentum component $\hat{L}_z$ is a conserved quantity characterized by the quantum number $m = 0, \pm 1, \pm 2,\dots$.
The wave function factorizes as $\Psi(\theta,\varphi) = \sum_{m\lambda} c_{m\lambda} \psi_m^{\lambda}(\theta) e^{i m \varphi}/\sqrt{2\pi}$ and, substituting this decomposition in the Schrödinger equation, we obtain
\begin{equation}
\bigg[ \hat{T}_{\theta} + \frac{m^2}{ 2 \rho^2(\theta)}
- \epsilon_{m}^{\lambda}  \bigg] \psi_m^{\lambda}(\theta) = 0,
\label{schreq0}
\end{equation}
with $\epsilon_m^{\lambda}$ the energy eigenvalue for a certain $m$ indexed by the real value ${\lambda}$ and normalization set to
\begin{equation}
\int_I  d\theta \, \rho(\theta) t(\theta) \, |\psi_m^{\lambda} (\theta)|^2=1.
\end{equation}
Note that the assumption of purely two-dimensional motion is applicable for energies $\epsilon_{m}^{\lambda}$ much smaller than the transverse confinement energy on the surface $\Sigma$.

The ground-state solution of Eq.~\eqref{schreq0} has zero quantum numbers $m=\lambda=0$ and is a nodeless constant function corresponding to zero energy:
\begin{equation}
\bar{\psi}_0^{0}(\theta) = \sqrt{2\pi/S}, \quad \bar{\epsilon}_{0}^0 = 0,
\end{equation}
so that the two-dimensional ground-state wave function reads $\bar{\Psi}_0(\theta,\varphi) = 1/\sqrt{S}$ (given that $c_{m\lambda} = \delta_{m0} \delta_{\lambda 0}$)
\footnote{The ground-state solution is uniform because we assume a uniform gas confinement throughout the surface.
Trapping inhomogeneities producing effective two-dimensional external potentials can be easily included in the Schrödinger equation.}.
All other real solutions of Eq.~\eqref{schreq0} constitute the excited-state components $\psi_m^{\lambda}(\theta)$ and the corresponding spectrum $\epsilon_{m}^{\lambda}$ of a quantum particle constrained to move on $\Sigma$.
Note that, for any value of the angular momentum projection $m$, there are infinite solutions labeled by the real quantum number $\lambda$.
These can be obtained numerically for specified choices of $\rho(\theta)$, $z(\theta)$, and $m$.
In the next sections, in particular, we will solve the problem in the ellipsoidal and toroidal cases.

\subsection{Ellipsoidal surface}
We parametrize the ellipsoid of semi-axes $a$ and $b$ by $\rho(\theta) = a \sin \theta$ and $z(\theta) = b \cos \theta$, where $\theta \in I = [0,\pi]$ {(see bottom of Fig.~\ref{fig1})}.
Substituting these formulas in Eq.~\eqref{schreq0}, we obtain
\begin{equation}
\left[- \frac{\partial_{\theta}^2}{2t^2(\theta)}  -\frac{a^2\cot\theta}{2t^4(\theta)} \partial_{\theta} + \frac{m^2}{2 \rho^2(\theta)} - \frac{\lambda(\lambda+1)}{2 a^2}  \right]\psi_m^{\lambda}(\theta) = 0,
\label{schreq}
\end{equation}
where $t(\theta) = (a^2 \cos^2\theta + b^2 \sin^2 \theta)^{1/2}$ and we redefined the energy as $\epsilon_{m}^{\lambda} = \lambda(\lambda+1)/(2 a^2)$ to introduce the real quantum number $\lambda$.

Note that Eq.~\eqref{schreq} depends only on the ratio $b/a$ between the semi-axes.
In particular, the ellipsoid is oblate for $b/a<1$, is prolate for $b/a>1$, and reduces to a sphere for $b/a=1$. 
Before proceeding further, we review the spherical case, whose Schrödinger equation reduces to
\begin{equation}
\left[ \frac{\hat{L}^2}{2 a^2} - \frac{l(l+1)}{2 a^2}  \right]\psi_m^{l}(\theta) = 0,
\end{equation}
with $\hat{L}^2 = - \partial_{\theta}^2 -\cot\theta \, \partial_{\theta} + m^2/\sin\theta^2$ the angular momentum operator and $\lambda \equiv l = 0,1,2,3,...
$ the corresponding integer quantum number.
The wave function can be written explicitly as  $\psi_m^{l}(\theta) = \sqrt{2\pi/a^2} \mathcal{Y}_m^{l}(\theta,0)$ in terms of the spherical harmonics $\mathcal{Y}_m^{l}(\theta,\varphi)$, with eigenenergies being degenerate in $m$. 

\begin{figure}[hbtp]
\centering
\includegraphics[width=0.99\columnwidth]{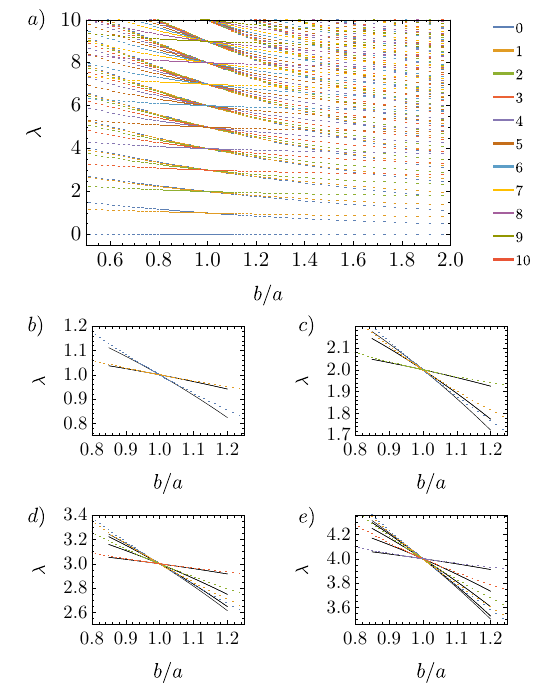}
\caption{a) Single-particle spectrum on the ellipsoidal surface, represented in terms of the quantum number $\lambda$, versus the aspect ratio $b/a$.
Note that $\lambda$ is the ellipsoidal-case analogous of the total angular momentum of a particle on the sphere, and breaking the spherical symmetry removes the level degeneracy in $m$.
The values of $|m|$ are indicated in the legend.
(b)-(e) Magnifications of (a) around sets of increasing values of $\lambda$. The black solid lines show the prediction of Eq.~(\ref{Eq:linearized}), obtained with first-order perturbation theory in the small parameter $e^2=(1-b^2/a^2)$. 
}
\label{fig2}
\end{figure}

Let us now consider the general case of an ellipsoid. 
The ground state of Eq.~\eqref{schreq} has zero energy $\bar{\epsilon}_{0}^{0}=0$ and corresponds to a flat nodeless solution $\bar{\psi}_0^0(\theta) = \sqrt{2\pi/S}$, where $S = 2 \pi a^2 [1+(1-e^2) \text{arctanh}(e)/e ]$ is the area of the ellipsoid and $e^2 = 1-b^2/a^2$ is the eccentricity.
Since no analytical solution is known for the excited states, we numerically solve Eq.~\eqref{schreq} to find the energy levels as a function of the ratio $b/a$. 
The obtained results for the quantum number $\lambda$ are presented in Fig.~\ref{fig2}{, while the eigenfunctions up to $m=2$ are shown in Appendix \ref{appB}}.
In the spherical case $\lambda$ assumes the integer values $0,1,2,3,...$ and can be interpreted as the quantum number of total angular momentum, with degenerate eigenenergies corresponding to different $m$. 
Such degeneracy is lifted in the ellipsoidal case, and in particular the values of $\lambda$ are shifted up in the oblate case ($b<a$), while they are shifted down in the prolate case ($b>a$). 
Note that the shift is maximal for the $m=0$ state and decreases in magnitude for higher $|m|$ values. 

We also develop a perturbation theory to evaluate the energy shift with respect to the spherical case.
In particular, we expand the Schrödinger equation \eqref{schreq} to first order in the small parameter $e^2=(1-b^2/a^2)$, obtaining
\begin{equation}
\left[ \frac{\hat{L}^2 -e^2 \hat{L}^{2'}}{2a^2} - \epsilon_{m}^{\lambda}  \right] \psi_m^{\lambda}(\theta)= 0,
\label{pertell}
\end{equation}
where $\hat{L}^{2'} = \sin^2\theta \partial_{\theta}^2 + \sin (2\theta) \partial_{\theta}$, and the linear-order expressions for the wave function and the energy are given by $\psi_m^{\lambda}(\theta) = \psi_m^{l}(\theta) -e^2  \psi_m^{\lambda'}(\theta)$ and $\epsilon_{m}^{\lambda} = \epsilon_{m}^{l} -e^2  \epsilon_{m}^{\lambda'}$. 
The unperturbed $e^2 =0$ problem is solved by $\psi_m^{l}(\theta) = \sqrt{2\pi/a^2} \mathcal{Y}_m^{l}(\theta,0)$ and has energy $ \epsilon_{m}^{l} = l(l+1)/(2a^2)$, while the first-order correction to the $(l,m)$ state energy is obtained by projecting Eq.~\eqref{pertell} over the unperturbed wave functions and neglecting $(e^2)^2$ terms. 
This operation yields
\begin{align}
\begin{split}
\lambda = l -e^2  \, \frac{2\pi}{2l+1}  \int_0^{\pi} d\theta \sin \theta \, \mathcal{Y}_m^{l*}(\theta,0) \hat{L}^{2'} \mathcal{Y}_m^{l}(\theta,0).
\end{split}
\label{Eq:linearized}
\end{align}
We compare the linear-order result with the exact calculation in the bottom panels of Fig.~\ref{fig2}, finding good agreement.
Note that the linear-order formula can also be calculated analytically by using the recurrence properties of the associated Legendre polynomials \cite{eswarathasan2022laplace}.

\subsection{Toroidal surface}

The torus can be parametrized by great ($R$) and small ($r$) circle radii as $\rho(\theta) = R + r \cos \theta$ and $z(\theta) = r \sin \theta$, with $\theta \in I = [0,2\pi]$ {(see the bottom of Fig.~\ref{fig1})}.
Substituting this parametrization in Eq.~\eqref{schreq0} gives
\begin{equation}
\bigg[- \frac{\partial_{\theta}^2}{2 r^2} +  \frac{\sin\theta}{ 2r\rho(\theta)} \partial_{\theta} + \frac{m^2}{ 2\rho^2(\theta)} - \frac{\lambda^2}{2r^2}  \bigg] \psi_m^{\lambda}(\theta) = 0,
\label{schreqtorus}
\end{equation}
where we define $\lambda^2 = 2 r^2 \epsilon_m^{\lambda}$ and impose periodic boundary conditions $\psi_m^{\lambda}(\theta) = \psi_{m}^{\lambda}(\theta+2\pi)$.
Note that Eq.~\eqref{schreqtorus} only depends on the ratio of the radii $r/R$. 
This aspect ratio is the only geometric parameter characterizing the torus surface, which evolves from a minimal nonintersecting doughnut-shaped form ($r/R = 1$) to a long cylinderlike surface ($r/R \ll 1$).

Let us first solve the problem in the cylindrical limit of $r/R \to 0$, in which $\rho(\theta)/r \to \infty$. 
In this limit, the Schrödinger equation simplifies to $(- \partial_{\theta}^2 -{l}^2) \psi_m^{l}(\theta) = 0$ with $\lambda \equiv l = 0, \pm 1, \pm 2, ...$ labeling the angular momentum of the particle rotating along the small circle.
The analytical wave function is, in this case, given by $\psi_m^{l}(\theta) \propto e^{i l \theta} $ and the eigenenergies are $\epsilon_{m}^{l} = {l}^2/(2r^2)$.
Note that the dependence on $m$ disappears since, in the cylindrical limit, the energy scale proportional to $1/R^2$ associated with the rotation along the $z$ axis vanishes in front of the kinetic energy $\epsilon_{m}^{l} \propto 1/r^2$ along the $r$ ring. 

\begin{figure}[hbtp]
\centering
\includegraphics[width=0.98\columnwidth]{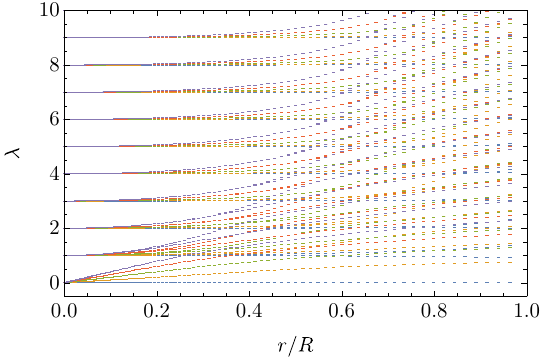}
\caption{
Single-particle spectrum $\lambda$ (related to the energy as $\epsilon_m^{\lambda}=\lambda^2/(2 r^2)$) on the torus surface versus the ratio $r/R$ for $|m| = 0,...,4$ (same colors as in Fig.~\ref{fig2}). 
In the limit $r/R \to 0$, the spectrum of a particle on a ring of radius $r$ is reproduced, which is doubly degenerate in $\pm l$ (see text). 
Outside the ring regime, these energies split.}
\label{fig3}
\end{figure}

We now solve the problem for arbitrary values of the major $R$ and minor $r$ radii.
The ground-state solution has energy $\bar{\epsilon}_0^{0}=0$ and wave function $\bar{\psi}_0^0(\theta) = \sqrt{2\pi/S}$, with $S = 4 \pi^2 Rr$ the torus area.
The solution of Eq.~\eqref{schreqtorus} for the excited states is obtained numerically and the resulting energy spectrum is reported in Fig.~\ref{fig3}, while the eigenfunctions up to $m=2$ are shown in Appendix \ref{appB}. 
In particular, for any values of $r/R$ and $m$, we obtain a ladder of excited states labeled by the real number $\lambda >0$.
These excited states are always separated by a gap from the ground-state energy.
We observe that, from $r/R \sim 0.5$ and as $r/R \to 0$, couples of adjacent $\lambda$ values corresponding to the same $m$ get closer and completely merge when $r/R = 0$. 
At $r/R=0$ these solutions correspond to degenerate states with quantum numbers $+l$ and $-l$. 
It is evident that the degeneracy of these levels is lifted by the curvature of the torus (see also Fig.~\ref{fig7} in Appendix \ref{appB}).

\section{Bose-Einstein condensation}

Let us now discuss how Bose-Einstein condensation is affected by the axially-symmetric geometry. 
We examine a gas of $N$ noninteracting bosons confined on the manifold $\Sigma$, assuming that the system is in thermal equilibrium at temperature $T$ and has a chemical potential $\mu$.
The total number of atoms can be expressed as
\begin{equation}
N = N_0 + N_T,
\label{NBEC}
\end{equation}
where the particle occupation numbers of the condensate and of the thermally excited states are given by
\begin{equation}
N_0 = \frac{1}{e^{(\bar{\epsilon}_{0}^0 -\mu)/ T}-1}, \quad N_T = \sum_{m\lambda} \frac{1}{e^{(\epsilon_m^{\lambda}-\mu)/ T}-1},
\label{N0NTBEC}
\end{equation}
and we set the Boltzmann constant to $k_B = 1$.

Although Bose-Einstein condensation does not occur at $T>0$ in infinite two-dimensional systems, it can still occur in finite-size ones.
In particular, an ideal Bose gas confined on the compact surface $\Sigma$ condenses in the single-particle ground state $\bar{\Psi}_0$ if its coherence length scale is larger than the system size. 
In terms of the chemical potential, the coherence criterion can be formulated as $|\mu| < \hbar^2/(2MS)$ \cite{hadzibabic2011two}, with $S$ the surface area.
We verify \textit{a posteriori} that this condition holds.

Let us define the temperature $T_{\text{BEC}}$, below which a fraction
of atoms start to occupy the condensate state significantly,
by setting  $\mu \approx \bar{\epsilon}_0^0 = 0$ in Eqs. \eqref{NBEC} and~\eqref{N0NTBEC} and assuming a fully depleted condensate $N_0 \to 0$. 
The resulting relation {$N= \sum_{m\lambda} (e^{\epsilon_m^{\lambda}/ T_{\text{BEC}}}-1)^{-1}$} can be evaluated numerically to get $T_{\text{BEC}}$ versus $N$ for a given geometry.
We present the results by rescaling the temperature as $\tilde{T}_{\text{BEC}} = k_B T / (\hbar^2 n /M)$, where $\hbar^2n/M$ corresponds to the critical temperature of an ideal Bose gas in a square flat box up to corrections scaling as $\log N$ \cite{tononi2024shell}.

\begin{figure}[t]
\centering
\includegraphics[width=0.98\columnwidth]{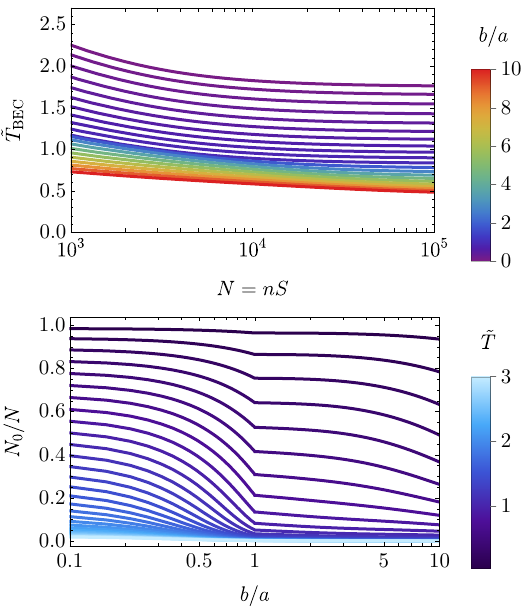}
\caption{Shown on top is the critical temperature $\tilde{T} = k_B T_{\text{BEC}} / (\hbar^2 n /M)$ versus $N=nS$ for different aspect ratios in an ellipsoid with $0.1 < b/a < 10$.
The colorbar indicates the value of $b/a$ and $r/R$ and in particular the curves correspond to nine equally spaced values of $b/a$ in $[0.1,0.9]$.
Shown on the bottom is the condensate fraction of a gas of $N=10^4$ particles confined on the ellipsoidal surface versus the aspect ratio for various temperature values $\tilde{T}$ (color bar).
For clarity, dimensionful units are reintroduced for $\tilde{T}$.
}
\label{fig4}
\end{figure}

\begin{figure}[t]
\centering
\includegraphics[width=0.98\columnwidth]{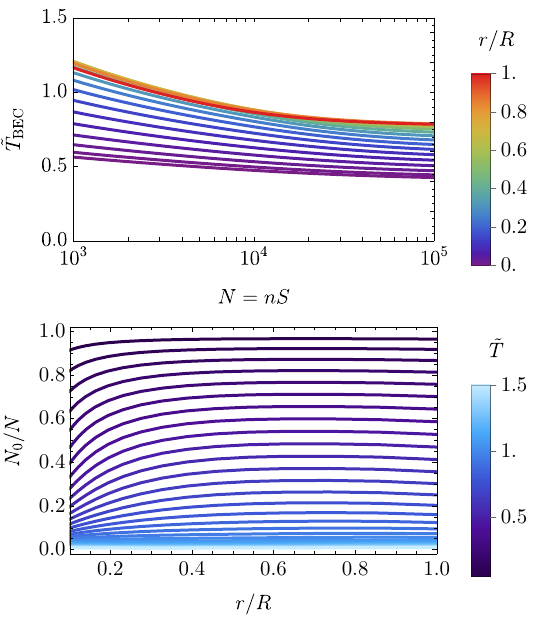}
\caption{Shown on top is the critical temperature $k_B T_{\text{BEC}} / (\hbar^2 n /M)$ versus $N=nS$ for different aspect ratios in a torus with $0.1 < r/R < 0.999$.
The colorbar indicates the value of $r/R$ and in particular the curves correspond to 16 values of $r/R$ distributed as $\sinh^2(r/R)$ in $[0.1,0.999]$.
Shown on the bottom is the condensate fraction of a gas of $N=10^4$ particles confined on the torus surface versus the aspect ratio for various temperature values $\tilde{T}$ (color bar).
Dimensionful units are reinserted in $\tilde{T}$ for clarity.
}
\label{fig5}
\end{figure}

Our results for the ellipsoid are shown in Fig.~\ref{fig4} and those for the torus in Fig.~\ref{fig5}.
The top panels depict $\tilde{T}_{\text{BEC}}$ versus $N$ for different aspect ratios.
We first note that, in agreement with the Mermin-Wagner theorem \cite{PhysRevLett.17.1133}, $\tilde{T}_{\text{BEC}}$ of both geometries tends logarithmically to zero when fixing the density $n$ and taking the thermodynamic limit $S \to \infty$.
Concerning the ellipsoid, we see that $\tilde{T}_{\text{BEC}}$ decreases as the aspect ratio $b/a$ is increased.
This geometric suppression of Bose-Einstein condensation is due to the change of geometry from a highly oblate pancake-shaped surface for $b/a \ll 1$ to a highly prolate cigar-shaped surface for $b/a \gg 1$.
The oblate geometry exhibits a more two-dimensional character, while the prolate geometry tends towards a one-dimensional manifold.
For the torus, similarly, we observe that Bose-Einstein condensation is disfavored as $r/R$ decreases.
In this case, the torus evolves from a doughnut-shaped surface with a two-dimensional character ($r/R=1$) to a long, thin cylinder with periodic boundaries ($r/R \ll 1$) exhibiting a one-dimensional behavior.
As a result, we also find that the specific eigenenergies distribution of the torus surface produces a slightly nonmonotonic behavior of $\tilde{T}_{\text{BEC}}$.
This subtle effect is ultimately determined by the energy-level distribution.

With our estimates of $\tilde{T}_{\text{BEC}}$, the residual condensate fraction at $\tilde{T} > \tilde{T}_{\text{BEC}}$ is neglected.
Going beyond this approximation requires the self-consistent evaluation of Eqs. \eqref{NBEC} and \eqref{N0NTBEC}, yielding the condensate fraction $N_0/N$ for a given number of atoms $N$, temperature $T$, and geometry. 
We calculate $N_0/N$ and show our results in the bottom panels of Figs.~\ref{fig4} and \ref{fig5}.

The ellipsoid condensate fraction vanishes quickly with temperature when $b/a \ll 1$, while it tends to zero more slowly when $b/a \gg 1$.
This qualitative difference in the behavior of $N_0/N$ agrees with the geometric suppression of $\tilde{T}_{\text{BEC}}$ in elongated surfaces.
Note that $N_0/N$ is not analytical around $b/a = 1$. 
At that point, the excited energy levels cross (see the bottom panels of Fig.~\ref{fig2}) 
and their occupations exhibit nonanalytic behavior in $b/a$, reflected in the discontinuity of the slope of $N_0/N$ at $b/a = 1$.
For the torus, similarly, the condensate fraction vanishes more slowly as $r/R \to 0$.
In addition, $N_0/N$ displays a faint maximum at intermediate values of $r/R$, which is what produces the non-monotonic behavior $\tilde{T}_{\text{BEC}}$.

The validity of our theory is verified by checking \textit{a posteriori} that the relation $|\mu| < \hbar^2/(2MS)$ \cite{hadzibabic2011two} holds in the regimes reported in the figures. Moreover, we calculate the flat-case $g_2$ correlation function at distances corresponding to the ellipsoid and torus sizes and find that it decays at temperatures $T \gtrsim T_{\text{BEC}}$.

\section{Bogoliubov spectrum on axially-symmetric surfaces}

Our formalization of the single-particle problem provides insight also into the many-body properties of the system. 
For instance, the formal solution of Eq.~\eqref{schreq0}, allows us to calculate the Bogoliubov excitation spectrum of a weakly repulsive bosonic gas confined on the axially-symmetric surface $\Sigma$. 
The Bogoliubov approach, inherently linked to the structure of the single-particle spectrum, accurately describes only weak interactions.
Beyond this regime, nonlinear phenomena such as swallow-tail structures \cite{PhysRevA.61.023402, PhysRevA.71.033622} of the excitation spectrum may emerge, potentially induced by the interplay of stronger interactions with periodic boundary conditions.

Let us describe the condensate via the time-dependent Gross-Pitaevskii equation (GPE) for the field $\Psi(\theta,\varphi,t)$,
\begin{equation}
i\partial_t \Psi(\theta,\varphi,t) = \left [ \hat{T} +g |\Psi(\theta,\varphi,t)|^2  \right ]\Psi(\theta,\varphi,t),
\label{2DGPE}
\end{equation}
where $g$ is the effective two-dimensional interaction strength.
This GPE can be derived via a dimensional reduction procedure from the mean-field action of a three-dimensional gas strongly confined near the surface $\Sigma$. 
In particular (see Refs. \cite{moller2020bose, biral2024bose, de2025geometric}), one assumes that the trapped condensate has a Gaussian profile perpendicular to the manifold $\Sigma$.
Equation~\eqref{2DGPE} is then obtained by assuming a uniform Gaussian thickness and neglecting the geometric potential \cite{moller2020bose, biral2024bose, de2025geometric}. 
The uniform thickness in particular allows us to assume a constant interaction strength $g$ throughout the surface, and its experimental implementation is thus necessary to ensure the applicability of our result.

The Bogoliubov spectrum is obtained by linearizing Eq.~\eqref{2DGPE} according to the standard Bogoliubov theory.
In particular, we expand the field $\Psi(\theta,\varphi,t)$ around the macroscopically occupied single-particle condensate state as $\Psi(\theta,\varphi,t) = [\sqrt{N} \bar{\Psi}_0 +\eta (\theta,\varphi,t)] e^{-i \mu t}$, where $\eta$ is a complex fluctuation field and the chemical potential is, at the lowest order, $\mu = g n_0$, with $n_0=N |\bar{\Psi}_0|^2 = N/S$.
By substituting this decomposition in the GPE and linearizing the result, we obtain
\begin{equation}
i \partial_t \eta(\theta,\varphi,t) = \left[ \hat{T} + 2gn_0 - \mu \right] \eta(\theta,\varphi,t) + gn_0 \eta^{*}(\theta,\varphi,t).
\label{fluctuationeq}
\end{equation}
We then expand the fluctuation field
\begin{equation}
\eta(\theta,\varphi,t) = u_m^{\lambda} \psi_m^{\lambda}(\theta) e^{i m \varphi} e^{i E_{m}^{\lambda} t} - v_m^{\lambda} \psi_m^{\lambda}(\theta) e^{-i m \varphi} e^{-i E_{m}^{\lambda} t}
\end{equation}
in terms of noninteracting Bogoliubov quasiparticles of energy $E_{m}^{\lambda}$ and substitute it in Eq.~\eqref{fluctuationeq}.
We note that eventual degeneracies in the single-particle levels (for instance the one in $\pm m$) will correspond to degeneracies of the Bogoliubov modes.
The analogous observation holds for a gas in a three-dimensional cubic box, where degenerate single-particle states labeled by momenta with equal magnitude but different directions yield the same Bogoliubov energy.
Finally, applying Eq.~\eqref{schreq0} and separating the resulting equation into negative- and positive-energy eigenmodes, we arrive at a system of Bogoliubov-de Gennes equations, which can be diagonalized to get the energy spectrum 
$E_{m}^{\lambda} = \sqrt{(\epsilon_m^{\lambda}+2gn_0-\mu)^2-(gn_0)^2}$. 
We substitute the lowest-order approximation of the chemical potential $\mu = g n_0$, obtaining the Bogoliubov spectrum
\begin{equation}
E_B = \sqrt{\epsilon_m^{\lambda} (\epsilon_m^{\lambda} + 2 g n_0)}.
\label{EBog}
\end{equation}
Note that, given the numerical single-particle energies for the axially symmetric surface $\Sigma$, one can order them in increasing order and obtain the Bogoliubov spectrum numerically for different values of the interaction strength.

In general, the distribution of the Bogoliubov energy modes can be categorized into two qualitatively different regimes of low and high energy.
The low-energy excitations, whose wavelength is comparable to either the local curvature radius or the system size are sensitive to the curved geometry. 
Instead, the high-energy excitations that correspond to wavelengths much smaller than both the local curvature radius and the system size, are not affected by the curvature. 
Their statistical distribution is similar to that of a gas in the two-dimensional flat geometry. 
This difference, depending on the specific choice of the axially symmetric surface $\Sigma$, can cause quantitative changes in the quantum statistical properties of the interacting system.

Let us now discuss our results in view of the applicability to ellipsoidal shell-shaped gases.
Interatomic interactions are expected to play a minor role in the currently available experimental regimes \cite{carollo2022observation}, but future experiments may be able to reach higher atomic densities and observe the Bogoliubov spectrum of Eq.~\eqref{EBog}.
In addition, interparticle interactions allow superfluidity, thus raising the question of how Bose-Einstein condensation interplays with the superfluid Berezinskii-Kosterlitz-Thouless transition in a curved finite-size geometry. 
This analysis, already conducted in the spherical case \cite{tononi2019bose}, is expected to be more complex in the cases of ellipsoids and other generic geometries due to the tensorial nature of the superfluid order parameter for surfaces of nonconstant curvature \cite{tononi2020quantum}. 
This interesting problem is left for future investigations.

\section{Conclusion}

We have studied the influence of the curved geometry on the energy spectrum in axially-symmetric surfaces, focusing on the experimentally relevant cases of an ellipsoid and a torus.
In particular, we formulated the one-body problem for a quantum particle confined on axially symmetric manifolds and applied it to both geometries.
We showed that, while the spectrum is degenerate in the specific limits of the sphere and the cylinder, the degeneracy is lifted in the general case. Therefore, the geometric parameters significantly influence the one-body physics of the system.
Furthermore, we analyzed ideal Bose-Einstein condensation, {discussing how the critical temperature and the condensate fraction are} affected by the geometric crossover between two-dimensional-like surfaces and elongated one-dimensional-like manifolds. 

Concerning possible applicability to the experiments, we emphasize that, while two-dimensional ellipsoidal shells were obtained in the thermal regime \cite{carollo2022observation}, the challenge of observing two-dimensional condensate shells is still open.
Our work addresses this regime, and can be easily extended to include one-body external potentials that model trapping inhomogeneities. 
The theory of ideal Bose-Einstein condensation is also useful for quantitative predictions, since interactions are expected to have a minor effect on the typical atom numbers of the experiments \cite{carollo2022observation}. 
Beyond this, our calculation of the Bogoliubov spectrum enables future investigations of the interplay between Bose-Einstein condensation and superfluidity in various curved geometries.
Finally, we mention that the one-body framework developed in this paper can be extended to address the few-body problem in generic axially symmetric surfaces.

\begin{acknowledgements}
A.T. acknowledges financial support from the Horizon Europe program HORIZON-CL4-2022-QUANTUM-02-SGA via the project 101113690 (PASQuanS2.1), {and funding by the European Union under the Horizon Europe MSCA programme via the project 101146753 (QUANTIFLAC).}
ICFO group acknowledges support from:
{European Research Council} AdG NOQIA;
MCIN/AEI ({PGC2018-0910.13039/ 501100011033}, {CEX2019-000910-S/10.13039/501100011033}, Plan National FIDEUA PID2019-106901GB-I00, Plan National STAMEENA PID2022-139099NB-I00, project funded by {MCIN/AEI/10.13039/501100011033} and by the ``{European Union NextGenerationEU/PRTR}'' ({PRTR-C17.I1}), FPI); QUANTERA DYNAMITE PCI2022-132919, QuantERA II Programme co-funded by {European Union's Horizon 2020 program} under Grant Agreement No {101017733};
{Ministry for Digital Transformation and of Civil Service of the Spanish Government} through the QUANTUM ENIA project call - Quantum Spain project, and by the {European Union} through the Recovery, Transformation and Resilience Plan - NextGenerationEU within the framework of the Digital Spain 2026 Agenda;
Fundaci\'o Cellex;
Fundaci\'o Mir-Puig;
Generalitat de Catalunya ({European Social Fund FEDER} and {CERCA program}); 
Barcelona Supercomputing Center MareNostrum (FI-2023-3-0024);
Funded by the European Union. Views and opinions expressed are however those of the author(s) only and do not necessarily reflect those of the European Union, European Commission, European Climate, Infrastructure and Environment Executive Agency (CINEA), or any other granting authority.  Neither the European Union nor any granting authority can be held responsible for them (HORIZON-CL4-2022-QUANTUM-02-SGA  PASQuanS2.1, 101113690, EU Horizon 2020 FET-OPEN OPTOlogic, Grant No 899794, QU-ATTO, 101168628),  {EU Horizon Europe Program} (This project has received funding from the {European Union's Horizon Europe research and innovation program} under grant agreement No {101080086} NeQSTGrant Agreement {101080086} --- NeQST);
ICFO Internal ``QuantumGaudi'' project;
G.E.A. further acknowledges the support of the Spanish Ministry of Science and Innovation (MCIN/AEI/10.13039/501100011033, grant PID2023-147469NB-C21), the Generalitat de Catalunya (grant 2021 SGR 01411) and {Barcelona Supercomputing Center MareNostrum} ({FI-2025-1-0020}).
\end{acknowledgements}

\appendix

\section{Evaluation of the kinetic energy operator}
\label{appA}
We express the kinetic energy operator in the coordinates $u=(\theta,\varphi)$ in the form $\hat{T} = -\Delta /2$. 
In particular, $\Delta$ is the Laplace-Beltrami operator, i.e. the Laplacian in the system of curved coordinates $u$. 
It is defined as
\begin{equation}
\Delta = \frac{1}{\sqrt{g}} \partial_i (\sqrt{g} g^{ij} \partial_j),
\end{equation}
where $\partial_i = \partial_{u_i}$, the metric tensor is defined as $g_{ij} = \partial_i \Sigma \cdot \partial_j \Sigma$ and its inverse as $g^{ij} = (g_{ij})^{-1}$, and the determinant $g=\det(g_{ij})$.

The Schrödinger equation \eqref{schreq} is obtained by evaluating explicitly the above operator for the chosen parametrization of the surface $\Sigma$.
In particular, the diagonal matrix $g_{ij}$ is given by 
\begin{equation}
g_{ij} = \begin{pmatrix}
t^2(\theta) & 0 \\
0 & \rho^2(\theta)
\end{pmatrix},
\end{equation}
so that $\sqrt{g} = \rho(\theta) t(\theta)$, and the surface area of the manifold is simply given by $S = \int_{0}^{2\pi} d \varphi \int_I  d\theta \, \sqrt{g}$, coinciding with Eq.~\eqref{surfacearea}.

\section{Single-particle eigenfunctions}
\label{appB}

\begin{figure}[hbtp]
\centering
\includegraphics[width=0.97\columnwidth]{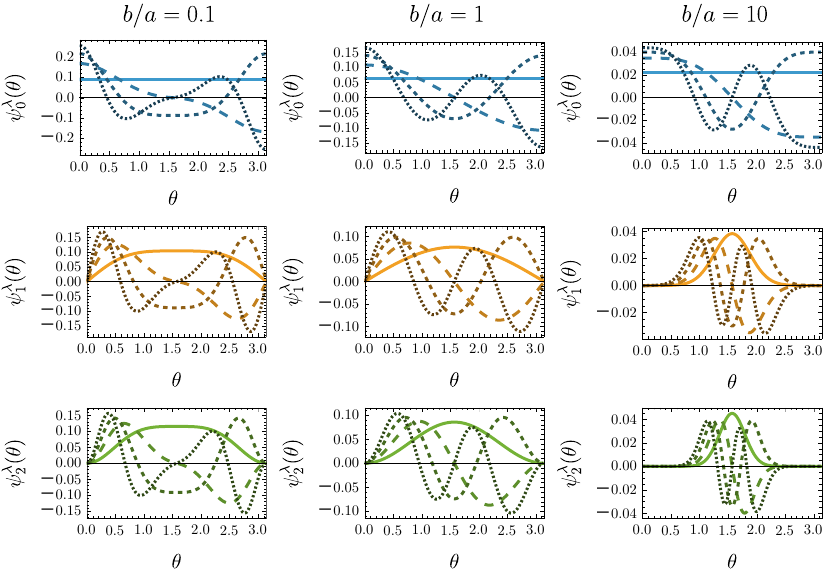}
\caption{
Eigenfunctions of a particle on the ellipsoidal surface.
From left to right, the values of $b/a$ correspond to a pancake-shaped ellipsoid, a sphere, and a cigar-shaped ellipsoid.
The rows correspond, from top to bottom, to the values of $m = \{ 0, 1, 2 \}$.
Note that the $m=0$ manifold includes the uniform ground state and the excited states with zero derivatives at the ellipsoidal poles $\theta = 0,\pi$, while all $m>0$ manifolds include eigenfunctions vanishing at the poles.
Increasingly excited states correspond to darker color shades and shorter dashes.
}
\label{fig6}
\end{figure}

\begin{figure}[hbtp]
\centering
\includegraphics[width=0.97\columnwidth]{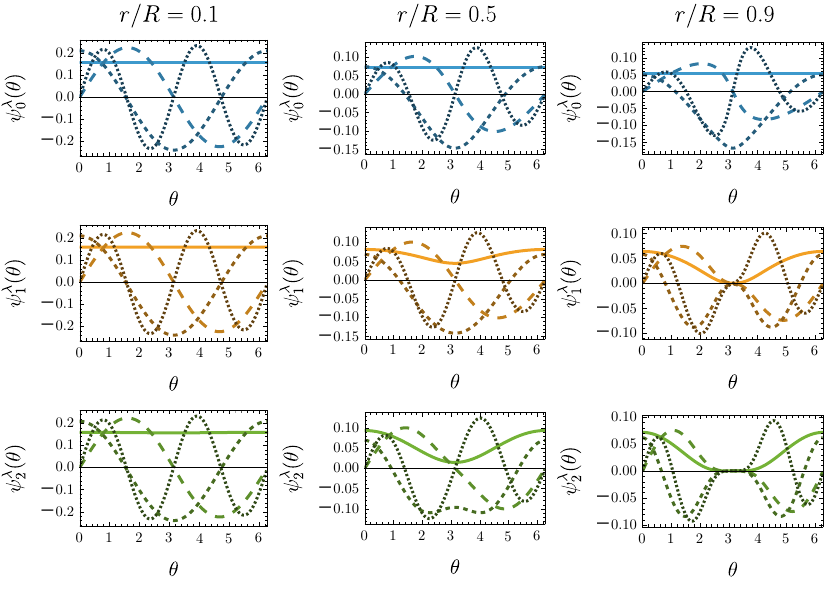}
\caption{
Eigenfunctions of a particle on the toroidal surface. From left to right, the values of $r/R$ span surfaces going from cylinderlike to doughnut shaped. 
The rows correspond, from top to bottom, to the values of $m = \{ 0, 1, 2 \}$.
In the cylinder limit ($r/R \to 0$) the eigenfunctions tend to harmonic functions with an increasing number of nodes. Increasingly excited states  correspond to darker color shades and shorter dashes.
}
\label{fig7}
\end{figure}

We report the eigenfunctions $\psi_m^{\lambda}(\theta)$ of the single-particle Schrodinger equation in Fig.~\ref{fig6} (ellipsoid) and in Fig.~\ref{fig7} (torus).
In particular, we fix the angular momentum projection along the $z$ axis to $m=0,1,2$ (first, second, and third rows) and show the four lowest-energy states for each $m$ value.
The color coding {of the lowest eigenstate} matches the one of Figs.~\ref{fig2} and \ref{fig3}. 
The excited state number for each $m$ value increases with darker color shades and shorter dashes.

The wave functions of particle on the ellipsoid spread throughout the surface when $b/a \ll 1$ (pancake-shaped ellipsoid). 
Instead, when $b/a \gg 1$ (cigar-shaped ellipsoid), the $m>0$ excited modes squeeze and tend to occupy the region around the ellipsoid equator $\theta = \pi/2$.

The wave functions of a particle on the torus can be interpreted in comparison with the cylinder case ($r/R = 0$), while outside this limit the same must follow by continuity.
We recall that the eigenfunctions for $r/R = 0$ take the $m$-independent form proportional to $\psi_m^{l}(\theta) \sim e^{i l \theta} $, with integer $l$ and energy degeneracy for equal $|l|$ values.
Degenerate eigenstates of the one-dimensional Schr\"odinger equation can be always combined to get real eigenfunctions \cite{Landau1991}, restituted by our numerical matrix-diagonalization routine.
In particular, linear combinations of the degenerate states for $l= \pm 1, \pm 2, \dots$ produce the series of harmonic functions $\sin\theta$, $\cos\theta$, $\sin (2\theta)$, $\cos(2\theta)$, $\dots$.
Although the degeneracy is lifted when $r/R >0$, this pattern is still recognizable for $r/R = 0.1$.
Indeed, on top of the uniform $l=0$ solution, we find eigenfunctions similar to the series of harmonic functions. 
These wave functions have, as expected, nodes near $|l|\theta = \pm \pi/2+2\pi c$ (for cosinelike eigenstates) or nodes near $|l|\theta = 2\pi c$ or $|l|\theta = \pi + 2\pi c$ (for sinelike eigenstates), with integer $c$.
When $r/R$ increases, the almost-degenerate energy levels further split and the differences in the corresponding eigenfunctions become more evident. 

\onecolumngrid

\end{document}